\documentclass{ecai} 

\usepackage{latexsym}
\usepackage{amssymb}
\usepackage{amsmath}
\usepackage{amsthm}
\usepackage{booktabs}
\usepackage{enumitem}
\usepackage{graphicx}
\usepackage{color}
\usepackage{booktabs}
\usepackage{multirow}
\usepackage{array}
\usepackage{makecell}
\usepackage{footnote}
\usepackage{tablefootnote}

\newcommand{\BibTeX}{B\kern-.05em{\sc i\kern-.025em b}\kern-.08em\TeX}

\begin{document}


\begin{frontmatter}


\paperid{123} 


\title{Leveraging LLMs for Scalable Non-intrusive \\ Speech Quality Assessment}


\author[A]{\fnms{Fredrik}~\snm{Cumlin}}
\author[B]{\fnms{Xinyu}~\snm{Liang}}
\author[A]{\fnms{Anubhab}~\snm{Ghosh}}
\author[A]{\fnms{Saikat}~\snm{Chatterjee}}

\address[A]{KTH Royal Institute of Technology, Sweden}
\address[B]{HCLTech, Sweden}


\begin{abstract}
Non-intrusive speech quality assessment (SQA) systems suffer from limited training data and costly human annotations, hindering their generalization to real-time conferencing calls. In this work, we propose leveraging large language models (LLMs) as pseudo-raters for speech quality to address these data bottlenecks. We construct \textit{LibriAugmented}, a dataset consisting of $101,129$ speech clips with simulated degradations labeled by a fine-tuned auditory LLM (Vicuna-7b-v1.5). We compare three training strategies: using human-labeled data, using LLM-labeled data, and a two-stage approach (pretraining on LLM labels, then fine-tuning on human labels), using both DNSMOS Pro and DeePMOS. We test on several datasets across languages and quality degradations. While LLM-labeled training yields mixed results compared to human-labeled training, we provide empirical evidence that the two-stage approach improves the generalization performance (e.g., DNSMOS Pro achieves 0.63 vs. 0.55 PCC on NISQA\_TEST\_LIVETALK and 0.73 vs. 0.65 PCC on Tencent with reverb). Our findings demonstrate the potential of using LLMs as scalable pseudo-raters for speech quality assessment, offering a cost-effective solution to the data limitation problem.

\end{abstract}

\end{frontmatter}


\section{Introduction}

Speech quality assessment (SQA) is the task of assessing the speech quality of speech clips. It plays an important role in modern communication systems, including voice-over-IP applications and speech enhancement algorithms \cite{hasqi, haaqi, dns_challenge2020, dns_challenge2021}. The gold standard in SQA is subjective evaluation using Mean Opinion Score (MOS), where human listeners rate speech quality on an ordinal scale from one (poor quality) to five (excellent quality) \cite{MOSbook}. However, subjective evaluations are costly and time-consuming, hence, there is a need for automated objective methods.

Non-intrusive SQA systems, which predict speech quality without reference signals, are particularly valuable for real-world applications where the reference audio is unavailable. Contemporary approaches suitable for applications in conferencing systems use deep neural networks (DNN) trained on subjectively rated datasets to map acoustic features (e.g., spectrogram, Mel-spectrogram) to MOS values \cite{DNSMOS, DNSMOSp, MOSNet, DeePMOS}. These models typically employ smaller architectures due to computational constraints in real-time speech quality monitoring applications. Despite several works having been introduced over the years, training non-intrusive SQA systems suffers from poor generalization to out-of-distribution data \cite{SSL-MOS, generalization}, a critical limitation in deployment as speech quality monitors in conferencing systems.

The generalization challenge stems largely from training data limitations, as discussed in \cite{LaMOSNet, MBNet, LDNet, DeePMOS, DeePMOS-B}. There are two main limitations: Not enough coverage of distortions and poor reliability of human ratings. For example, VCC2018 \cite{VCC2018} and BVCC \cite{BVCC} contain only synthetic speech distortions, and TMHINT-QI \cite{tmhintqi} contains only additive background noise and enhanced clips thereof. The NISQA Corpus \cite{NISQA} contains a large variety of distortions, but to the best of the authors' knowledge, it is still lacking some very common communication artifacts such as high reverberant scenarios, echo leakage, and different speech attenuation distortions. Mixing of existing datasets is difficult due to biases in subjective SQA; a subjective evaluation depend on the context of the evaluation (contextual bias \cite{contextual_bias}), the distribution of the quality of the clips included in the evaluation (range-equalizing bias \cite{range-equalizing}), and the granularity of the evaluation scale \cite{koster_bias}. Furthermore, the open-sourced dataset typically only contains 3-8 human ratings per clip, producing large confidence intervals \cite{MOSNet}. Collecting comprehensive human-rated datasets across diverse conditions remains prohibitively expensive.

In this work, we propose using large language models (LLMs) as pseudo-raters to address this data bottleneck. Recent auditory LLMs combining audio encoders (e.g., Whisper \cite{whisper}) with transformer decoders have shown strong performance on speech perception tasks \cite{fathullah2024prompting, yu2024connecting, zheng2024bat}. These models can process audio and perform quality assessments at scale, potentially covering degradation scenarios underrepresented in human-rated datasets. Our approach investigates two key questions. First, can training non-intrusive SQA models suitable for applications in conferencing systems on LLM-labeled speech improve generalization compared to human-labeled training? Second, does pretraining on LLM-labeled data, followed by fine-tuning on human-labeled data--a so-called two-stage training approach--enhance performance on out-of-distribution data?

We use a customized version of the LLM Vicuna-7b-v1.5 \cite{vicuna}, finetuned for speech quality assessment tasks \cite{wang2025enabling}, to generate quality labels for a large-scale dataset derived from LibriSpeech \cite{librispeech} with simulated degradations. This gives us a large dataset that is LLM-labeled. We then evaluate whether using this dataset as a training dataset for the smaller non-intrusive SQA systems improves the generalization performance. Our two-stage training framework then combines the large LLM-labeled dataset with the smaller human-labeled dataset to investigate the second research question.

Our contributions are as follows. First, we introduce a methodology for leveraging LLMs as pseudo-raters for speech quality, enabling scalable and cost-effective generation of labeled datasets for training, thereby addressing the first research question. Second, we propose a two-stage training framework that combines pretraining on LLM-labeled data with fine-tuning on human-rated data to improve generalization, addressing the second research question. And third, we provide empirical evidence that incorporating LLM-generated labels into the training process enhances the generalization performance of non-intrusive SQA systems.

\subsection{Relevant literature}

Non-intrusive SQA using DNNs has evolved significantly since 2016, where one of the earliest approaches mapped MFCC features of speech to quality values using convolutional neural networks (CNNs) and global max pooling \cite{yoshimura16_interspeech}. AutoMOS \cite{AutoMOS} and QualityNet \cite{QualityNet} continued by exploring LSTM architectures instead of the CNNs, and QualityNet used frame-level scores to improve stability at the time of training. MOSNet followed the QualityNet method but combined CNNs with LSTM, showing improved performance \cite{MOSNet}. This work was followed by several other works, but with different approaches to deal with limitations in the training data, as discussed in the introduction \cite{MBNet, LDNet, LaMOSNet, DeePMOS, DeePMOS-B}.

Recent developments have focused on probabilistic modeling \cite{DeePMOS, DeePMOS-B, DNSMOSp} and semi-supervised learning (SSL) approaches \cite{SSL-MOS, UTMOS, multivariate}. The SSL-based approaches use a large pretrained feature extractor (e.g., wav2vec 2.0 \cite{w2v2}) and fine-tune a prediction head on speech quality tasks. These models have shown good generalization performance on out-of-distribution datasets, and significantly outperform the end-to-end trained counterparts \cite{SSL-MOS, UTMOS}. However, the SSL-based approaches also have a significantly larger computation and memory cost than the smaller counterparts, making them ill-suited for monitoring quality in real-time communication systems \cite{DNSMOSp}.

To the authors' knowledge, the incorporation of LLMs in non-intrusive SQA is a new research direction, where some of the earliest papers come from 2024/2025 \cite{zezario2025study, chen2025audio, wang2025enabling}. Zezario et al. investigate zero-shot non-intrusive SQA by exploring GPT-4o's direct speech quality capabilities and GPT-4o's speech quality capability by providing a transcription of the speech (produced by an automatic speech recognizer), where the transcription serves as a proxy for the speech quality \cite{zezario2025study}. They empirically show that the zero-shot approaches outperform DNSMOS on a Chinese speech quality dataset. Chen et al. address the limitation that most audio LLMs remain unaware of the speech quality by introducing a speech evaluation corpus from a human-rated dataset, which enables audio LLMs to provide both MOS predictions and descriptive comparisons between speech samples \cite{chen2025audio}. Wang et al. focus on fine-tuning audio LLMs on speech quality data \cite{wang2025enabling}. These works have collectively started to set a foundation for using LLMs in speech quality assessment, though they primarily focus on direct evaluation rather than the pseudo-labeling approach explored in this work.

\section{Method}

Consider a subjectively rated speech dataset $\mathcal{D} = \{(\mathbf{x}_n, y_n)\}_{n=1}^N$, where $\mathbf{x}_n$ denotes the features extracted from the $n$-th speech clip and $y_n$ is the corresponding quality score. There are $N$ clips in total. The quality score $y_n$, typically in the MOS scale, is given by the average rating provided by humans. The number of ratings per speech clip typically varies between three to eight.

The goal of non-intrusive SQA is to learn a regressor function $f_\theta$, parameterized by $\theta$, that maps input features $\mathbf{x}_n$ to the quality score $y_n$. Since $f_\theta$ operates only on the potentially distorted speech signal without requiring a clean reference, it is suitable for deployment in real-world, reference-free scenarios. In practice, $f_\theta$ is often instantiated as a deep neural network and trained on $\mathcal{D}$ using stochastic gradient descent.

This work focuses on the generalization ability of the regressor $f_\theta$: its capacity to produce reliable quality predictions for speech clips that differ in content or distortion type from those seen during training. To evaluate generalization, assume we have a quality rated training data $\mathcal{D}_{\text{train}}$, along with several test datasets $\mathcal{D}_{\text{test}}^{(m)}$ for $m=1,..., M$, where $M$ is the number of test datasets. To test the generalizability of a regressor, we train the regressor on $\mathcal{D}_{train}$, and test on the test datasets $\mathcal{D}_{\text{test}}^{(m)}$ for $m=1,..., M$. The quality values on the test data are human-rated labels, since the purpose of a non-intrusive SQA system is to emulate subjective quality scores. The performance measure can be, for example, Pearson's correlation coefficient (PCC) or Spearman's rank correlation coefficient (SRCC), as per \cite{DNSMOSp}.

\subsection{LibriAugmented: $\mathcal{D}_{LLM}$}
\label{subsection:libriaugmented}

\begin{figure}[t]
  \centering
  \includegraphics[width=7cm]{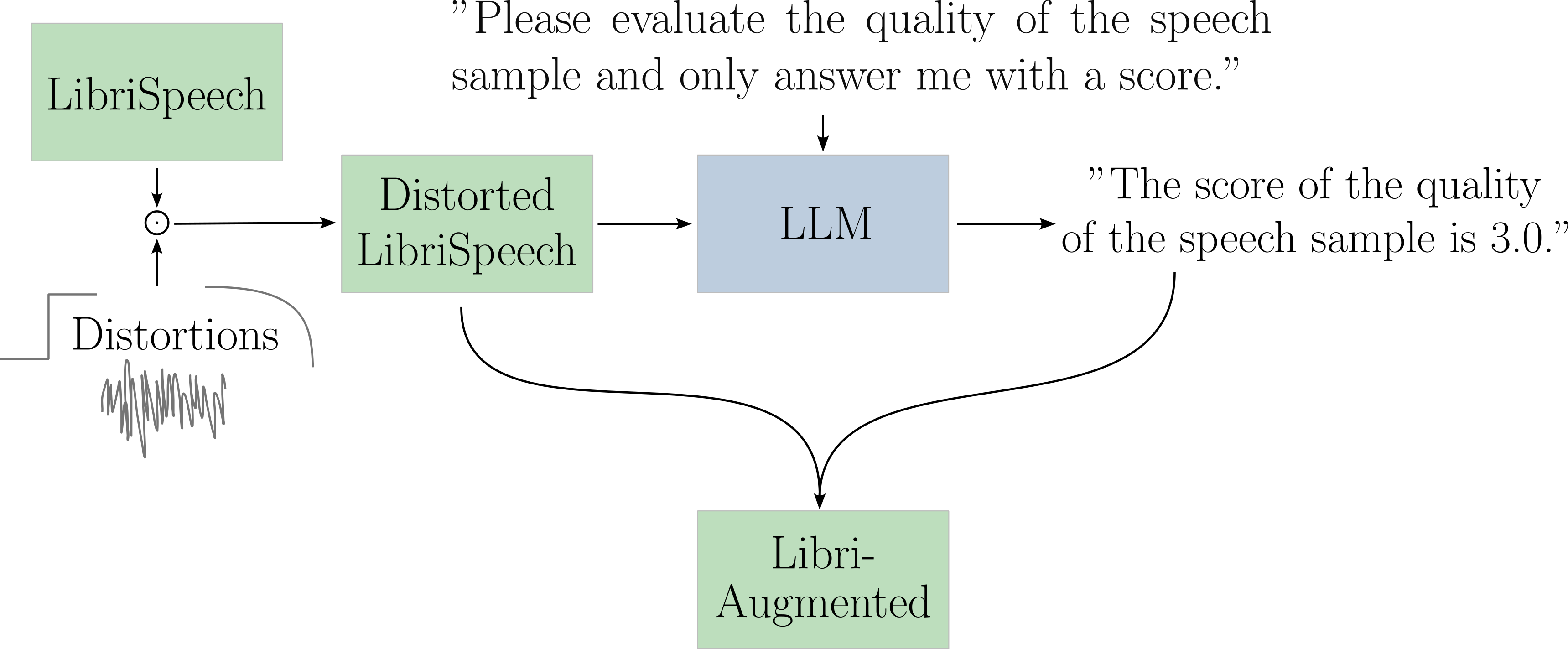}
  \caption{Creation of the LibriAugmented dataset.}
  \label{fig:libriaugmented}
\end{figure}

To address our research questions, whether training a non-intrusive SQA system on LLM-labeled data improves generalization, we construct a speech dataset with simulated distortions and label it with an LLM. The construction follows the construction of the LibriAugmented dataset design in \cite{impairments}. We start with clean speech samples from the LibriSpeech \cite{librispeech} dataset, and we simulate diverse distortion scenarios using the audiomentations library\footnote{\url{https://github.com/iver56/audiomentations}}, applying 15 different types of distortions, as well as a clean condition (i.e., no distortion), to mimic realistic impairments that may occur in online conferencing environments. This gives a distorted LibriSpeech dataset, totaling $101,129$ speech clips. The distribution of distortion types in the dataset is shown in Table~\ref{tab:libriaugmented}. 


We label the distorted LibriSpeech dataset with an adopted state-of-the-art (SOTA) auditory LLM fine-tuned for MOS prediction \cite{wang2025enabling}. The LLM in question is Vicuna-7b-v1.5 \cite{vicuna}. The adopted LLM incorporates a Whisper encoder with a Low-Rank Adaptation (LoRA) \cite{hu2022lora}, with the latter being finetuned to a speech quality task. Each speech clip was labeled by the LLM by providing the clip together with the prompt "\textit{Please evaluate the quality of the speech sample and only answer me with a score}". This gives an LLM-labeled speech dataset, which we call LibriAugmented. A schematic overview of the creation of LibriAugmented is given in Fig.~\ref{fig:libriaugmented}.


\begin{table}[ht!]
\caption{The LibriAugmented dataset. $\mathcal{D}_{LLM}$}

\begin{center}
\begin{tabular}{c c}
\hline
\rule{0pt}{3ex}
\textbf{Ratio} & \textbf{Impairment (Parameters)} \\[2ex]
\hline
\rule{0pt}{3ex}

$0.050$ & Identity \\[1.5ex]
$0.050$ & AddBackgroundNoise (snr=$-10\sim 15$ dB) \\[1.5ex]
$0.050$ & ClippingImpairment (percentile=$10\sim40\%$) \\[1.5ex]
$0.050$ & GainTransition (gain=$-60\sim20$ dB) \\[1.5ex]
$0.050$ & LowPassFilter (cutoff\_freq=$0.5\sim1$ kHz) \\[1.5ex]
$0.050$ & Mp3Compression (bit\_rate=$8\sim14$) \\[1.5ex]
$0.050$ & PitchShift (semitones=$-4\sim4$ kHz) \\[1.5ex]
$0.050$ & RoomSimulator (rt60=$0.8\sim1.5$ s) \\[1.5ex]
$0.050$ & TimeMask (band\_part=$0.2\sim0.5$) \\[1.5ex]
$0.050$ & TimeStretch (rate=$0.5\sim2$) \\[2ex]


$0.083$ & AddBackgroundNoise + RoomSimulator \\[1.5ex]
$0.083$ & AddBackgroundNoise + LowPassFilter \\[1.5ex]
$0.083$ & AddBackgroundNoise + TimeStretch \\[1.5ex]
$0.083$ & RoomSimulator + Mp3Compression \\[1.5ex]
$0.083$ & PitchShift + LowPassFilter \\[1.5ex]
$0.083$ & GainTransition + TimeMask \\[1.5ex]

\hline
\end{tabular}
\end{center}

\label{tab:libriaugmented}
\end{table}

\subsection{Training and pre-training using $\mathcal{D}_{LLM}$}

Using $\mathcal{D}_{LLM}$, we will do two experiments.
First, we will train a non-intrusive SQA model $f_\theta$ on the LibriAugmented dataset $\mathcal{D}_{LLM}$ and then test on the standard human-labeled datasets $\mathcal{D}_{test}^{(m)}$ for $m=1,.., M$. We then compare this to the performance of the same SQA model, but when trained on a human-labeled dataset. This addresses the first research question, whether training on an LLM-labeled speech dataset improves generalization compared to training on a human-labeled dataset.

\begin{figure}[t]
  \centering
  \includegraphics[width=8cm]{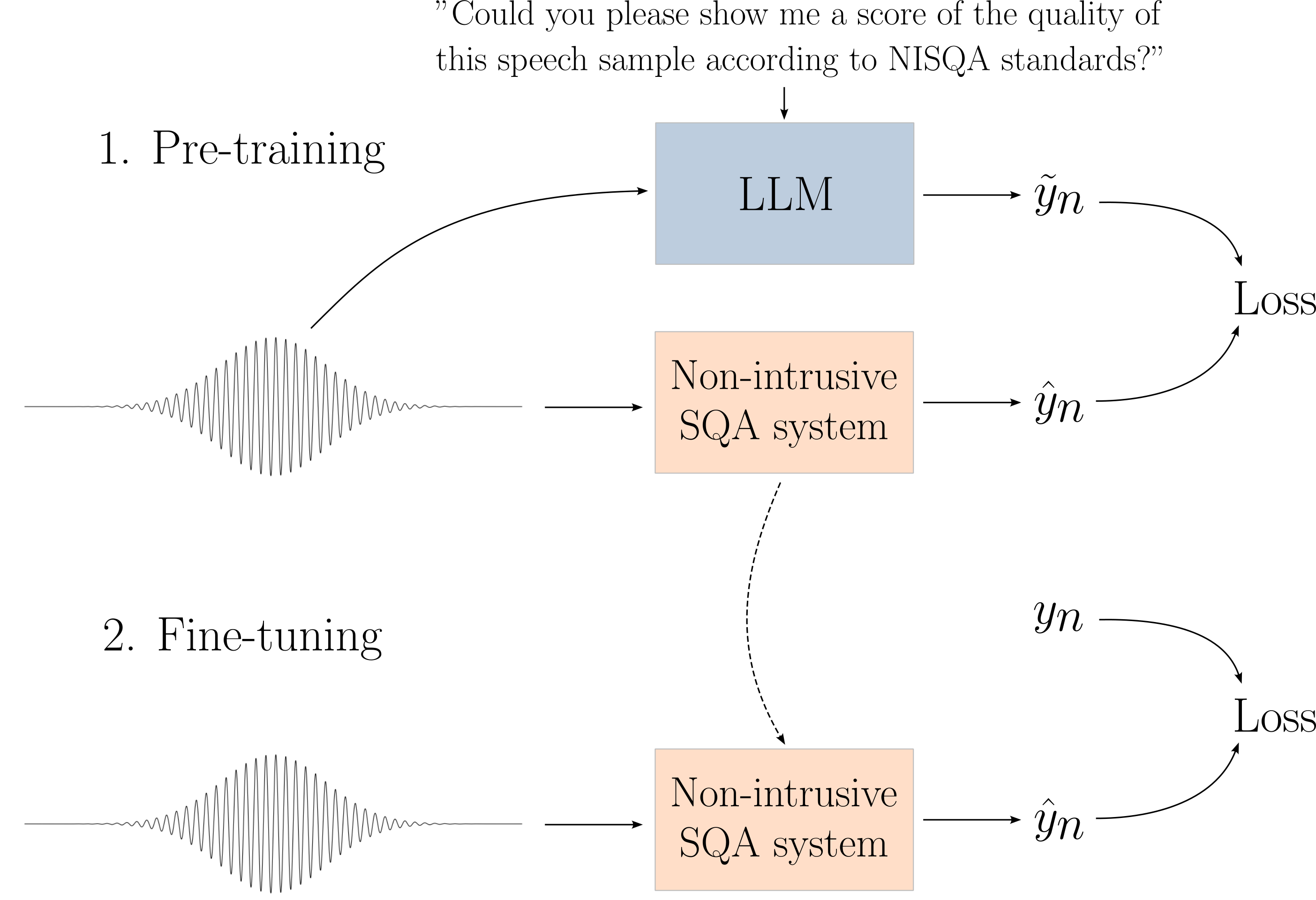}
  \caption{Overview of pre-training a non-intrusive SQA model on LLM-labeled data, followed by fine-tuning on human-labeled data.}
  \label{fig:libriaugmented_pretraining}
\end{figure}

Second, we will first pre-train a non-intrusive SQA model $f_\theta$ on the LibriAugmented dataset $\mathcal{D}_{LLM}$. This is followed by a fine-tuning stage on a human-labeled dataset, and then tested on standard human-labeled datasets. Comparing this to the same SQA model when trained only on human-labeled data, we address our second research question, whether pre-training on LLM-labeled data followed by fine-tuning on human-labeled data improves generalization. An overview of this training procedure is given in Fig.~\ref{fig:libriaugmented_pretraining}.

\section{Experiments}

\begin{table*}[ht]
\centering
\caption{Speech Quality Assessment Datasets}
\label{tab:speech_datasets}
\footnotesize
\setlength{\tabcolsep}{6pt}
\renewcommand{\arraystretch}{1.3}

\begin{tabular}{l|c|c|c|c|p{4cm}}
\toprule
\multirow{2}{*}{\textbf{Dataset}} & \multirow{2}{*}{\makecell{\textbf{\# Clips}}} & \multirow{2}{*}{\makecell{\textbf{\# Ratings}\\\textbf{per Clip}}} & \multirow{2}{*}{\textbf{Split}} & \multirow{2}{*}{\textbf{Language}} & \multirow{2}{*}{\textbf{Key Characteristics}} \\
& & & & & \\
\midrule
\multicolumn{6}{c}{\textbf{NISQA Corpus} \cite{NISQA}} \\
\midrule
NISQA\_TRAIN\_SIM & $10,000$ & $\sim5$ & train & English & Simulated degradations \\
NISQA\_VAL\_SIM & $2,500$ & $\sim5$ & val & English & Simulated degradations \\
NISQA\_TEST\_LIVETALK & $232$ & $24$ & test & German & Live conversational speech \\
NISQA\_TEST\_FOR & $240$ & $\sim30$ & test & (Australian) English & Simulated degradations+live \\
NISQA\_TEST\_P501 & $240$ & $\sim 28$ & test & (British) English & Simulated degradations+live \\
\midrule
\multicolumn{6}{c}{\textbf{TMHINT-QI} \cite{tmhintqi}} \\
\midrule
TMHINT-QI train & $10,000$ & $\sim 3$ & train & Chinese (Mandarin) & Simulated degradations \\
TMHINT-QI val & $2,937$ & $\sim 3$ & val & Chinese (Mandarin) & Simulated degradations \\
TMHINT-QI test & $1,978$ & $\sim 3$ & test & Chinese (Mandarin) & Simulated degradations \\
\midrule
\multicolumn{6}{c}{\textbf{Tencent Corpus} \cite{Tencent}} \\
\midrule
Tencent w/ reverb & $3,197$ & $\sim 28$ & test & Chinese (Mandarin) & Simulated degradations+live \\
Tencent w/o reverb & $6,914$ & $\sim 28$ & test & Chinese (Mandarin) & Simulated degradations \\
\midrule
\multicolumn{6}{c}{\textbf{LibriAugmented}} \\
\midrule
LibriAugmented train & $100,000$ & LLM rated & train & English & Simulated degradations \\
LibriAugmented val & $1,129$ & LLM rated & val & English & Simulated degradations \\
\bottomrule
\end{tabular}
\end{table*}

\subsection{Datasets}

We employ three datasets for training, three for validation, and six for testing. These datasets are selected from the NISQA Corpus \cite{NISQA}, TMHINT-QI \cite{tmhintqi}, the Tencent Corpus \cite{Tencent}, and our constructed LibriAugmented dataset (see Subsection \ref{subsection:libriaugmented}). The first three datasets are human-rated, and the last is LLM-rated. An overview of the datasets used in this work is provided in Table~\ref{tab:speech_datasets}.

\subsubsection{NISQA Corpus (human-rated)}

We utilize five datasets from the NISQA Corpus, each rated following the absolute category rating (ACR) scale in accordance with the ITU-T Recommendation P.808 \cite{p808}. Ratings are provided along five perceptual dimensions; however, we only consider the overall quality dimension, corresponding to the standard Mean Opinion Score (MOS).

NISQA\_TRAIN\_SIM is a simulated dataset consisting of $10,000$ clean speech clips that have been artificially degraded by applying various distortions such as Gaussian noise, signal correlated MNRU (modulated noise reference unit) noise, and filters with random cutoff frequencies. Each clip is rated by approximately five raters. This dataset is used for training.

NISQA\_VAL\_SIM is constructed analogously to NISQA\_TRAIN\_SIM, consisting of $2,500$ clips with similar distortion types and the same number of raters per clip. It serves as the validation set for model selection.

NISQA\_TEST\_LIVETALK is a live-recorded dataset containing $232$ German-language speech clips captured directly from terminal devices such as smartphones and laptops. It contains 'natural' disturbances such as babble noise (also known as cocktail noise), poor reception, busy street, etc. Each clip has 24 ratings. This dataset is used for testing.

NISQA\_TEST\_FOR is a simulated test set containing $230$ English (Australian-accented) clips degraded by common artifacts such as codec compression, background noise, and packet loss. Each clip is rated by approximately 30 raters. This dataset is used for testing.

NISQA\_TEST\_P501 is similar in construction to NISQA\_TEST\_FOR and contains $240$ English clips with around $28$ ratings per clip. This dataset is used for testing.

\subsubsection{TMHINT-QI (human-rated)}

The TMHINT-QI (Taiwan Mandarin Hearing in Noise Test - Quality Intelligibility) dataset is a simulated dataset constructed from clean Mandarin utterances that were distorted by additive background noises at varying SNR levels. A portion of the speech clips was processed by speech enhancement models after the addition of background noise. There is a predefined split in the dataset into train and test. The train set contains $12,937$ speech clips, and the test set contains $1,978$ speech clips. All clips are rated on the MOS scale with an average of approximately three ratings per clip.

We further split the train set into two; one that will be used as a training dataset and one that will be used for validation (model selection). We call the training dataset TMHINT-QI train, and this contains $10,000$ clips. We call the validation dataset TMHINT-QI val, and it contains $2,937$ clips. We call the original test dataset TMHINT-QI test.

\subsubsection{Tencent Corpus (human-rated)}

The Tencent Corpus comprises two datasets, Tencent with reverberation (Tencent w/ reverb) and Tencent without reverberation (Tencent w/o reverb). Tencent w/ reverb contains $3,197$ Mandarin utterances recorded or simulated with reverberation. These clips represent both real and synthetic degradation conditions and are rated by approximately 28 raters per clip. Tencent w/o reverb consists of $6,914$ Mandarin clips\footnote{This is a subset of the original dataset. Some of the clips were corrupted, hence, we only use the non-corrupted clips. In the original description of the dataset, there are ``about $10000$ Chinese speech clips'' in the w/o reverb dataset \cite{Tencent}.} with simulated distortions typical of online conferencing environments, but without added reverberation. It is also rated with the same number of human judgments per clip. Both datasets are used exclusively for testing.

\subsubsection{LibriAugmented (LLM-rated)}

A detailed description of the LibriAugmented dataset can be found in Subsection~\ref{subsection:libriaugmented}. Briefly, this dataset is derived from clean LibriSpeech utterances and augmented with diverse distortion types using \texttt{audiomentations}. Speech quality scores are generated using a fine-tuned auditory LLM. We divide the dataset into a training set of $100,000$ clips and a validation set of $1,129$ clips. This dataset is not used for any testing purposes.

\subsection{Non-intrusive SQA systems under consideration}

The models under consideration for the experiments are DNSMOS Pro \cite{DNSMOSp} and DeePMOS \cite{DeePMOS}. These two contemporary non-intrusive speech quality models are suitable since they are designed for real-time communication systems, only require the quality value as a target label, and operate at a sampling rate of $16$ kHz.

\subsubsection{DNSMOS Pro}

\begin{table*}[t]
  \centering
  \caption{PCC performance of DNSMOS Pro predictions on several test datasets, with different training datasets/strategies. For each training dataset/strategy, $10$ models are trained independently, and the mean $\pm$ standard deviation results are reported.}
  \label{tab:dnsmos_pro}
  \renewcommand{\arraystretch}{1.15}
  \begin{tabular}{@{}lccc ccc@{}}
    \toprule
    \multicolumn{1}{c}{\textbf{Test data →}} 
      & \multicolumn{3}{c}{\textbf{NISQA}} 
      & \textbf{TMHINT-QI} 
      & \multicolumn{2}{c@{}}{\textbf{Tencent}} \\
    \cmidrule(lr){2-4} \cmidrule(lr){5-5} \cmidrule(lr){6-7}
    \multicolumn{1}{c}{\textbf{Training data ↓}} & LIVETALK & FOR & P501 
      & test 
      & w/ reverb & w/o reverb \\
    \midrule
    NISQA\_TRAIN\_SIM   &  $0.55\pm 0.05$  &  $\mathbf{0.76}\pm 0.03$   &  $\mathbf{0.81}\pm 0.02$   &  $0.49\pm 0.04$   &    $0.65\pm 0.06$   &   $\mathbf{0.75}\pm 0.01$   \\
    TMHINT-QI  &  $0.12\pm 0.08$  &  $0.30\pm 0.04$   &  $0.39\pm 0.02$   &  $\mathbf{0.75}\pm 0.01$   &    $0.42\pm 0.10$   &   $0.56\pm 0.03$   \\
    LibriAugmented   &  $0.46\pm 0.07$  &  $0.37\pm 0.03$   &  $0.60\pm 0.03$   &  $0.46\pm 0.02$   &    $0.60\pm 0.03$   &   $0.70\pm 0.02$    \\
    LibriAug. + NISQA\_TRAIN\_SIM (finetune)  &  $\mathbf{0.63}\pm 0.01$  &  $0.73\pm 0.01$   &  $\mathbf{0.81}\pm 0.01$   &  $0.56\pm 0.01$   &    $\mathbf{0.73}\pm 0.01$   &   $0.71\pm 0.0$         \\
    \bottomrule
  \end{tabular}
\end{table*}

\begin{table*}[t]
  \centering
  \caption{PCC performance of DeePMOS predictions on several test datasets, with different training datasets/strategies. For each training dataset/strategy, $10$ models are trained independently, and the mean $\pm$ standard deviation results are reported.}
  \label{tab:deepmos}
  \renewcommand{\arraystretch}{1.15}
  \begin{tabular}{@{}lccc ccc@{}}
    \toprule
    \multicolumn{1}{c}{\textbf{Test data →}} 
      & \multicolumn{3}{c}{\textbf{NISQA}} 
      & \textbf{TMHINT-QI} 
      & \multicolumn{2}{c@{}}{\textbf{Tencent}} \\
    \cmidrule(lr){2-4} \cmidrule(lr){5-5} \cmidrule(lr){6-7}
    \multicolumn{1}{c}{\textbf{Training data ↓}} & LIVETALK & FOR & P501 
      & test 
      & w/ reverb & w/o reverb \\
    \midrule
    NISQA\_TRAIN\_SIM   &  $0.35\pm 0.09$  &  $0.68\pm 0.05$   &  $0.72\pm 0.04$   &  $0.41\pm 0.05$   &    $\mathbf{0.54}\pm 0.06$   &   $\mathbf{0.65}\pm 0.03$   \\
    TMHINT-QI  &  $0.08\pm 0.08$  &  $0.17\pm 0.01$   &  $0.29\pm 0.02$   &  $\mathbf{0.72}\pm 0.01$   &    $0.37\pm 0.09$   &   $0.31\pm 0.08$   \\
    LibriAugmented   &  $0.37\pm 0.11$  &  $0.46\pm 0.03$   &  $0.54\pm 0.04$   &  $0.38\pm 0.02$   &    $0.45\pm 0.07$   &   $0.59\pm 0.03$    \\
    LibriAug. + NISQA\_TRAIN\_SIM (finetune)  &  $\mathbf{0.54}\pm 0.03$  &  $\mathbf{0.77}\pm 0.01$   &  $\mathbf{0.80}\pm 0.02$   &  $0.41\pm 0.01$   &    $\mathbf{0.54}\pm 0.02$   &   $\mathbf{0.65}\pm 0.01$         \\
    \bottomrule
  \end{tabular}
\end{table*}

DNSMOS Pro is a variant of the DNSMOS architecture and training setup \cite{DNSMOS, DNSMOSp}. The input is the log-magnitude spectrogram of a (distorted) speech clip, and the output is a Gaussian distribution of the quality score. Thus, the model incorporates an estimation of the uncertainty in the quality prediction.

The architecture consists of an encoder, a global max pooling layer, and a prediction head. The encoder maps the spectrogram to a $3$-dimensional representation using a combination of $2$-dimensional convolutions and pooling layers, batch normalization, and dropout. The dropout rate is $0.3$. The global max pooling layer takes the maximum value along the channel dimension, resulting in a $1$-dimensional representation, which is then mapped down by the prediction head to a vector of $2$ elements representing a Gaussian distribution.

Training of DNSMOS Pro is done by minimizing the Gaussian negative log-likelihood using the quality label as target. It is trained using an Adam optimizer, batch size of $64$, and learning rate of $10^{-4}$. It is trained for 500 epochs, with the checkpoint achieving the highest correlation on the validation set selected as the final model.

\subsubsection{DeePMOS}

DeePMOS is an extension of the MOSNet architectural family \cite{MOSNet, DeePMOS}. Similar to DNSMOS Pro, it predicts a Gaussian distribution over the quality value given the magnitude spectrogram of a (distorted) speech clip. 

The architecture consists of an encoder, a bi-directional LSTM (bLSTM) layer, and a prediction head. The encoder consists of $2$-dimensional convolutions, ReLU activation functions, and dropout (of $0.3$). There are no pooling layers. Instead, every third layer, a stride of $(1,3)$ is used, meaning that the elements in the frequency dimension are reduced by a third, while the number of elements in the time dimension remains constant. This is followed by a bLSTM and then a dense layer, which predicts a distribution of the speech quality for each time bin of the spectrogram.

Training of DeePMOS is done by minimizing the Gaussian negative log-likelihood with the quality label as target for each time bin. This is called a frame-level loss \cite{QualityNet}. DeePMOS is using the same configuration of optimizer and batch size as DNSMOS Pro. It is trained for $60$ epochs, with the checkpoint achieving the highest correlation on the validation dataset selected as the final model.

\subsection{Results}

In this section we will discuss the results. The evaluation measure used is the Pearson's Correlation Coefficient (PCC), and it is given by the following. Let $\mathcal{D}=\{(\mathbf{x}_n, y_n)\}_{n=1}^{N}$ denote a dataset where $\mathbf{x}_n$ is speech clip features, $y_n$ is the corresponding MOS or LLM value (depending on dataset used), and $N$ is the number of clips. Let $\hat{y}_n$ denote a prediction of the quality value produced by a non-intrusive system. The PCC of the quality labels and the predictions is given by \begin{equation}
    \begin{split}
    PCC =  \frac{\sum_{n=1}^{N}(y_n-\overline{y})(\hat{y}_n-\overline{\hat{y}})}{\sqrt{\sum_{n=1}^{N}(y_n-\overline{y})^2}. \sqrt{\sum_{n=1}^{N}(\hat{y}_n-\overline{\hat{y}})^2}},
    \end{split}
\end{equation}

where $\overline{y}$ is the average value of the quality labels, and $\overline{\hat{y}}$ is the average value of the predictions. 

We implement DNSMOS Pro and DeePMOS and train according to their respective method in four different experimental settings:
\begin{enumerate}
    \item Training on NISQA\_TRAIN\_SIM, validating on NISQA\_VAL\_SIM.
    \item Training on TMHINT-QI train, validating on TMHINT-QI val.
    \item Training on LibriAugmented train, validating on LibriAugmented val.
    \item Pretraining on LibriAugmented train, validating on LibriAugmented val, and then fine-tuning on NISQA\_TRAIN\_SIM for $10$ epochs with no dropout in the architectures, validating on NISQA\_VAL\_SIM.
\end{enumerate}

For all experimental settings, we test on all the test datasets and report the PCC performance. The purpose of the validation datasets is for model selection; validation occurs at every epoch, and the model that achieves the highest PCC on the validation dataset is extracted. 

Furthermore, for each experimental setting, we implement $10$ models to train independently, for statistical consideration. The results for DNSMOS Pro can be seen in Table \ref{tab:dnsmos_pro}, and the results for DeePMOS can be seen in Table \ref{tab:deepmos}.

\textbf{Addressing the first research question:} Our first research question examined whether training non-intrusive SQA models on LLM-labeled data improves generalization compared to human-labeled training. The results demonstrate mixed outcomes for this approach. For both DNSMOS Pro and DeePMOS, when comparing the models trained solely on LibriAugmented (LLM-labeled) data to NISQA\_TRAIN\_SIM (human-labeled) data, the LLM-trained models generally underperformed on most test datasets. However, comparing the LLM-trained models to the models trained on TMHINT-QI train (human-labeled) data, the former outperformed on all but one test dataset. This result is consistent for both models. This suggests that the diverse synthetic degradations in the LibriAugmented dataset provide better coverage of distortions than the TMHINT-QI dataset, which is more important than having a subjectively rated dataset.

\textbf{Addressing the Second Research Question:} The second research question investigated whether a two-stage training approach—pretraining on LLM-labeled data followed by fine-tuning on human-labeled data—improves the generalization ability. The experimental results provide evidence for this. For DNSMOS Pro, models trained according to the two-stage approach achieved higher PCC on three test datasets, the same on one, and slightly worse on two, compared to models trained on NISQA\_TRAIN\_SIM. For DeePMOS the results are clearer; the trained models, according to the two-stage approach, achieved higher PCC on three test datasets, and the same on the other three, compared to the models trained on NISQA\_TRAIN\_SIM. This suggests that pre-training on the large-scale LLM-labeled data provides a robust foundation for subsequent fine-tuning, which improves generalization.

\textbf{Limitations:} We would also like to highlight some limitations with this experimental setup. First, the LibriAugmented dataset is a large dataset, but could be more varied. Only an English Corpus, the LibriSpeech Corpus, was used as clean speech clips, and simulated degradations were applied on top. An interesting extension is to include more languages--in particular Chinese and German for these test datasets--and include 'real' speech clips; that is, speech clips that do not have simulated degradation applied as a postprocessing step, instead, they are introduced at the time of recording. Furthermore, the finetuning stage could benefit from a more continual learning approach, incorporating more datasets than just one. 

\section{Conclusion}

In this work, we propose using LLM as a pseudo-rater to provide quality scores on speech clips. We created a dataset, called the LibriAugmented dataset, that contains synthetic distortions and quality ratings provided by an auditory LLM. We investigated two research questions: does training on LibriAugmented improve generalization? And does a two-stage approach--pre-training on LibriAugmented followed by fine-tuning on human-rated data--improve generalization? Our experimental results indicate mixed results for the first question and evidence that a two-stage approach improves generalization. This work has demonstrated the potential of using LLMs as scalable pseudo-raters for speech quality assessment. Future work is to explore a more diverse dataset to be rated by an LLM and explore other fine-tuning strategies.

\begin{ack}
This work was partially supported by the Wallenberg AI, Autonomous Systems and Software Program (WASP) funded by the Knut and Alice Wallenberg Foundation.

The computations and data handling were enabled by resources provided by Chalmers e-Commons at Chalmers.
\end{ack}


\bibliography{mybibfile}

\begin{thebibliography}{41}
\providecommand{\natexlab}[1]{#1}
\providecommand{\url}[1]{\texttt{#1}}
\expandafter\ifx\csname urlstyle\endcsname\relax
  \providecommand{\doi}[1]{doi: #1}\else
  \providecommand{\doi}{doi: \begingroup \urlstyle{rm}\Url}\fi

\bibitem[Baevski et~al.(2020)Baevski, Zhou, Mohamed, and Auli]{w2v2}
A.~Baevski, Y.~Zhou, A.~Mohamed, and M.~Auli.
\newblock wav2vec 2.0: A framework for self-supervised learning of speech
  representations.
\newblock \emph{Advances in neural information processing systems},
  33:\penalty0 12449--12460, 2020.

\bibitem[Chen et~al.(2025)Chen, Hu, Wang, Wang, Chen, Zhang, Yang, and
  Chng]{chen2025audio}
C.~Chen, Y.~Hu, S.~Wang, H.~Wang, Z.~Chen, C.~Zhang, C.-H.~H. Yang, and
  E.~Chng.
\newblock Audio large language models can be descriptive speech quality
  evaluators.
\newblock In \emph{The Thirteenth International Conference on Learning
  Representations}, 2025.

\bibitem[Chen et~al.(2022)Chen, Fu, Yu, Wang, and Tsao]{tmhintqi}
Y.-W. Chen, S.-W. Fu, C.~Yu, H.-M. Wang, and Y.~Tsao.
\newblock {InQSS: a speech intelligibility and quality assessment model using a
  multi-task learning network}.
\newblock In \emph{Proc. Interspeech 2022}, pages 146--150, 2022.
\newblock \doi{10.21437/Interspeech.2022-10064}.

\bibitem[Cooper and Yamagishi(2023)]{range-equalizing}
E.~Cooper and J.~Yamagishi.
\newblock Investigating range-equalizing bias in mean opinion score ratings of
  synthesized speech.
\newblock In \emph{Interspeech 2023}, pages 1104--1108, 2023.
\newblock \doi{10.21437/Interspeech.2023-1076}.

\bibitem[Cooper et~al.(2022)Cooper, Huang, Toda, and Yamagishi]{SSL-MOS}
E.~Cooper, W.-C. Huang, T.~Toda, and J.~Yamagishi.
\newblock Generalization ability of mos prediction networks.
\newblock In \emph{ICASSP 2022-2022 IEEE International Conference on Acoustics,
  Speech and Signal Processing (ICASSP)}, pages 8442--8446. IEEE, 2022.

\bibitem[Cumlin et~al.(2023)Cumlin, Sch\"uldt, and Chatterjee]{LaMOSNet}
F.~Cumlin, C.~Sch\"uldt, and S.~Chatterjee.
\newblock Latent-based neural net for non-intrusive speech quality assessment.
\newblock In \emph{2023 33th European Signal Processing Conference (EUSIPCO)},
  European Signal Processing Conference, pages 36--40, sep 2023.

\bibitem[Cumlin et~al.(2024{\natexlab{a}})Cumlin, Liang, and
  Chatterjee]{generalization}
F.~Cumlin, X.~Liang, and S.~Chatterjee.
\newblock Generalization ability of end-to-end non-intrusive speech quality
  models.
\newblock In \emph{2024 IEEE 21st India Council International Conference
  (INDICON)}, pages 1--5, 2024{\natexlab{a}}.
\newblock \doi{10.1109/INDICON63790.2024.10958371}.

\bibitem[Cumlin et~al.(2024{\natexlab{b}})Cumlin, Liang, Ungureanu, KA~Reddy,
  Sch{\"u}ldt, and Chatterjee]{DNSMOSp}
F.~Cumlin, X.~Liang, V.~Ungureanu, C.~KA~Reddy, C.~Sch{\"u}ldt, and
  S.~Chatterjee.
\newblock Dnsmos pro: A reduced-size dnn for probabilistic mos of speech.
\newblock In \emph{Proc. Interspeech 2024}, pages 4818--4822,
  2024{\natexlab{b}}.

\bibitem[Cumlin et~al.(2025{\natexlab{a}})Cumlin, Liang, Ungureanu, Reddy,
  Sch{\"u}ldt, and Chatterjee]{impairments}
F.~Cumlin, X.~Liang, V.~Ungureanu, C.~K. Reddy, C.~Sch{\"u}ldt, and
  S.~Chatterjee.
\newblock Impairments are clustered in latents of deep neural network-based
  speech quality models.
\newblock In \emph{ICASSP 2025-2025 IEEE International Conference on Acoustics,
  Speech and Signal Processing (ICASSP)}, pages 1--5. IEEE, 2025{\natexlab{a}}.

\bibitem[Cumlin et~al.(2025{\natexlab{b}})Cumlin, Liang, Ungureanu, Reddy,
  Sch{\"u}ldt, and Chatterjee]{multivariate}
F.~Cumlin, X.~Liang, V.~Ungureanu, C.~K. Reddy, C.~Sch{\"u}ldt, and
  S.~Chatterjee.
\newblock Multivariate probabilistic assessment of speech quality.
\newblock In \emph{Proc. Interspeech 2025}, 2025{\natexlab{b}}.

\bibitem[Fathullah et~al.(2024)Fathullah, Wu, Lakomkin, Jia, Shangguan, Li,
  Guo, Xiong, Mahadeokar, Kalinli, et~al.]{fathullah2024prompting}
Y.~Fathullah, C.~Wu, E.~Lakomkin, J.~Jia, Y.~Shangguan, K.~Li, J.~Guo,
  W.~Xiong, J.~Mahadeokar, O.~Kalinli, et~al.
\newblock Prompting large language models with speech recognition abilities.
\newblock In \emph{ICASSP 2024-2024 IEEE International Conference on Acoustics,
  Speech and Signal Processing (ICASSP)}, pages 13351--13355. IEEE, 2024.

\bibitem[Fu et~al.(2018)Fu, Tsao, Hwang, and Wang]{QualityNet}
S.~Fu, Y.~Tsao, H.~Hwang, and H.~Wang.
\newblock {Quality-Net: A}n end-to-end non-intrusive speech quality assessment
  model based on {BLSTM}.
\newblock In B.~Yegnanarayana, editor, \emph{Interspeech 2018, 19th Annual
  Conference of the International Speech Communication Association, Hyderabad,
  India, 2-6 September 2018}. {ISCA}, 2018.
\newblock \doi{10.21437/Interspeech.2018-1802}.

\bibitem[Hu et~al.(2022)Hu, Shen, Wallis, Allen-Zhu, Li, Wang, Wang, Chen,
  et~al.]{hu2022lora}
E.~J. Hu, Y.~Shen, P.~Wallis, Z.~Allen-Zhu, Y.~Li, S.~Wang, L.~Wang, W.~Chen,
  et~al.
\newblock Lora: Low-rank adaptation of large language models.
\newblock \emph{ICLR}, 1\penalty0 (2):\penalty0 3, 2022.

\bibitem[Huang et~al.(2022{\natexlab{a}})Huang, Cooper, Tsao, Wang, Toda, and
  Yamagishi]{BVCC}
W.-C. Huang, E.~Cooper, Y.~Tsao, H.-M. Wang, T.~Toda, and J.~Yamagishi.
\newblock The voicemos challenge 2022.
\newblock \emph{arXiv preprint arXiv:2203.11389}, 2022{\natexlab{a}}.

\bibitem[Huang et~al.(2022{\natexlab{b}})Huang, Cooper, Yamagishi, and
  Toda]{LDNet}
W.-C. Huang, E.~Cooper, J.~Yamagishi, and T.~Toda.
\newblock {LDNet: U}nified listener dependent modeling in {MOS} prediction for
  synthetic speech.
\newblock In \emph{ICASSP 2022 - 2022 IEEE International Conference on
  Acoustics, Speech and Signal Processing (ICASSP)}, 2022{\natexlab{b}}.
\newblock \doi{10.1109/ICASSP43922.2022.9747222}.

\bibitem[Kates and Arehart(2010)]{hasqi}
J.~M. Kates and K.~H. Arehart.
\newblock The hearing-aid speech quality index ({HASQI}).
\newblock \emph{Journal of the Audio Engineering Society}, 58\penalty0
  (6):\penalty0 454--463, June 2010.

\bibitem[Kates and Arehart(2016)]{haaqi}
J.~M. Kates and K.~H. Arehart.
\newblock The hearing-aid audio quality index (haaqi).
\newblock \emph{IEEE/ACM Transactions on Audio, Speech, and Language
  Processing}, 24\penalty0 (2):\penalty0 354--365, 2016.
\newblock \doi{10.1109/TASLP.2015.2507858}.

\bibitem[Köster et~al.(2015)Köster, Guse, Wältermann, and
  Möller]{koster_bias}
F.~Köster, D.~Guse, M.~Wältermann, and S.~Möller.
\newblock Comparison between the discrete acr scale and an extended continuous
  scale for the quality assessment of transmitted speech.
\newblock In \emph{Forschritte der Akustik – DAGA’15, 41. Jahrestagung für
  Akustik}, pages 150--152, Nürnberg, Germany, Mar. 2015.
\newblock Conference: DAGA 2015, Nürnberg.

\bibitem[Leng et~al.(2021)Leng, Tan, Zhao, Soong, Li, and Qin]{MBNet}
Y.~Leng, X.~Tan, S.~Zhao, F.~K. Soong, X.~Li, and T.~Qin.
\newblock {MBNet:} {MOS} prediction for synthesized speech with mean-bias
  network.
\newblock In \emph{{IEEE} International Conference on Acoustics, Speech and
  Signal Processing, {ICASSP} 2021, Toronto, ON, Canada, June 6-11, 2021}.
  {IEEE}, 2021.
\newblock \doi{10.1109/ICASSP39728.2021.9413877}.

\bibitem[Liang et~al.(2023)Liang, Cumlin, Sch\"uldt, and Chatterjee]{DeePMOS}
X.~Liang, F.~Cumlin, C.~Sch\"uldt, and S.~Chatterjee.
\newblock Deepmos: Deep posterior mean-opinion-score of speech.
\newblock In \emph{Interspeech 2023}. {ISCA}, Aug 2023.

\bibitem[Liang et~al.(2024)Liang, Cumlin, Ungureanu, Reddy, Sch{\"u}ldt, and
  Chatterjee]{DeePMOS-B}
X.~Liang, F.~Cumlin, V.~Ungureanu, C.~K. Reddy, C.~Sch{\"u}ldt, and
  S.~Chatterjee.
\newblock Deepmos-b: Deep posterior mean-opinion-score using beta distribution.
\newblock In \emph{2024 32nd European Signal Processing Conference (EUSIPCO)},
  pages 416--420. IEEE, 2024.

\bibitem[Lin et~al.(2011)Lin, Tao, Kacprzyk, Li, Izquierdo, and Wang]{MOSbook}
W.~Lin, D.~Tao, J.~Kacprzyk, Z.~Li, E.~Izquierdo, and H.~Wang.
\newblock \emph{Multimedia Analysis, Processing and Communications}.
\newblock Springer Publishing, New York, 2011.

\bibitem[Lo et~al.(2019)Lo, Fu, Huang, Wang, Yamagishi, Tsao, and Wang]{MOSNet}
C.-C. Lo, S.-W. Fu, W.-C. Huang, X.~Wang, J.~Yamagishi, Y.~Tsao, and H.-m.
  Wang.
\newblock {MOSNet: D}eep learning-based objective assessment for voice
  conversion.
\newblock In \emph{Interspeech 2019}, pages 1541--1545, 09 2019.
\newblock \doi{10.21437/Interspeech.2019-2003}.

\bibitem[Lorenzo-Trueba et~al.(2018)Lorenzo-Trueba, Yamagishi, Toda, Saito,
  Villavicencio, Kinnunen, and Ling]{VCC2018}
J.~Lorenzo-Trueba, J.~Yamagishi, T.~Toda, D.~Saito, F.~Villavicencio,
  T.~Kinnunen, and Z.~Ling.
\newblock The voice conversion challenge 2018: Promoting development of
  parallel and nonparallel methods, 2018.

\bibitem[Mittag et~al.(2021)Mittag, Naderi, Chehadi, and Möller]{NISQA}
G.~Mittag, B.~Naderi, A.~Chehadi, and S.~Möller.
\newblock {NISQA}: A deep {CNN}-self-attention model for multidimensional
  speech quality prediction with crowdsourced datasets.
\newblock In \emph{Interspeech 2021}. {ISCA}, Aug 2021.
\newblock \doi{10.21437/interspeech.2021-299}.

\bibitem[P.808(2018)]{p808}
I.-T.~R. P.808.
\newblock Subjective evaluation of speech quality with a crowdsourcing
  approach., 2018.

\bibitem[Panayotov et~al.(2015)Panayotov, Chen, Povey, and
  Khudanpur]{librispeech}
V.~Panayotov, G.~Chen, D.~Povey, and S.~Khudanpur.
\newblock Librispeech: An asr corpus based on public domain audio books.
\newblock In \emph{2015 IEEE International Conference on Acoustics, Speech and
  Signal Processing (ICASSP)}, pages 5206--5210, 2015.
\newblock \doi{10.1109/ICASSP.2015.7178964}.

\bibitem[Patton et~al.(2016)Patton, Agiomyrgiannakis, Terry, Wilson, Saurous,
  and Sculley]{AutoMOS}
B.~Patton, Y.~Agiomyrgiannakis, M.~Terry, K.~Wilson, R.~A. Saurous, and
  D.~Sculley.
\newblock {AutoMOS: L}earning a non-intrusive assessor of
  naturalness-of-speech.
\newblock In \emph{NIPS 2016 End-to-end Learning for Speech and Audio
  Processing Workshop}, 2016.

\bibitem[Radford et~al.(2023)Radford, Kim, Xu, Brockman, McLeavey, and
  Sutskever]{whisper}
A.~Radford, J.~W. Kim, T.~Xu, G.~Brockman, C.~McLeavey, and I.~Sutskever.
\newblock Robust speech recognition via large‑scale weak supervision.
\newblock In \emph{Proceedings of the 40th International Conference on Machine
  Learning}, volume 202 of \emph{Proceedings of Machine Learning Research},
  pages 28492--28518, Baltimore, MD, USA, July 2023. PMLR.

\bibitem[Reddy et~al.(2021{\natexlab{a}})Reddy, Gopal, and Cutler]{DNSMOS}
C.~Reddy, V.~Gopal, and R.~Cutler.
\newblock {DNSMOS: A} non-intrusive perceptual objective speech quality metric
  to evaluate noise suppressors.
\newblock In \emph{{IEEE} International Conference on Acoustics, Speech and
  Signal Processing, {ICASSP} 2021, Toronto, ON, Canada, June 6-11, 2021}, 06
  2021{\natexlab{a}}.
\newblock \doi{10.1109/ICASSP39728.2021.9414878}.

\bibitem[Reddy et~al.(2021{\natexlab{b}})Reddy, Dubey, Gopal, Cutler, Braun,
  Gamper, Aichner, and Srinivasan]{dns_challenge2021}
C.~K.~A. Reddy, H.~Dubey, V.~Gopal, R.~Cutler, S.~Braun, H.~Gamper, R.~Aichner,
  and S.~Srinivasan.
\newblock Icassp 2021 deep noise suppression challenge.
\newblock In \emph{ICASSP 2021 - 2021 IEEE International Conference on
  Acoustics, Speech and Signal Processing (ICASSP)}, pages 6623--6627,
  2021{\natexlab{b}}.
\newblock \doi{10.1109/ICASSP39728.2021.9415105}.

\bibitem[Saeki et~al.(2022)Saeki, Xin, Nakata, Koriyama, Takamichi, and
  Saruwatari]{UTMOS}
T.~Saeki, D.~Xin, W.~Nakata, T.~Koriyama, S.~Takamichi, and H.~Saruwatari.
\newblock Utmos: Utokyo-sarulab system for voicemos challenge 2022.
\newblock In \emph{Proc. Interspeech 2022}, pages 4521--4525, 09 2022.
\newblock \doi{10.21437/Interspeech.2022-439}.

\bibitem[Strake et~al.(2020)Strake, Defraene, Fluyt, Tirry, and
  Fingscheidt]{dns_challenge2020}
M.~Strake, B.~Defraene, K.~Fluyt, W.~Tirry, and T.~Fingscheidt.
\newblock Interspeech 2020 deep noise suppression challenge: A fully
  convolutional recurrent network (fcrn) for joint dereverberation and
  denoising.
\newblock In \emph{Interspeech 2020}, pages 2467--2471, 2020.
\newblock \doi{10.21437/Interspeech.2020-2439}.

\bibitem[Wang et~al.(2024)Wang, Éva Székely, and Gustafson]{contextual_bias}
S.~Wang, Éva Székely, and J.~Gustafson.
\newblock Contextual interactive evaluation of tts models in dialogue systems.
\newblock In \emph{Interspeech 2024}, pages 2965--2969, 2024.
\newblock \doi{10.21437/Interspeech.2024-1008}.

\bibitem[Wang et~al.(2025)Wang, Yu, Yang, Tang, Li, Zhuang, Chen, Tian, Zhang,
  Sun, et~al.]{wang2025enabling}
S.~Wang, W.~Yu, Y.~Yang, C.~Tang, Y.~Li, J.~Zhuang, X.~Chen, X.~Tian, J.~Zhang,
  G.~Sun, et~al.
\newblock Enabling auditory large language models for automatic speech quality
  evaluation.
\newblock In \emph{ICASSP 2025-2025 IEEE International Conference on Acoustics,
  Speech and Signal Processing (ICASSP)}, pages 1--5. IEEE, 2025.

\bibitem[Yi et~al.(2022)Yi, Xiao, Xiao, Naderi, Moller, Wardah, Mittag, Cutler,
  Zhang, Williamson, Chen, Yang, and Shang]{Tencent}
G.~Yi, W.~Xiao, Y.~Xiao, B.~Naderi, S.~Moller, W.~Wardah, G.~Mittag, R.~Cutler,
  Z.~Zhang, D.~S. Williamson, F.~Chen, F.~Yang, and S.~Shang.
\newblock Conferencingspeech 2022 challenge: Non-intrusive objective speech
  quality assessment (nisqa) challenge for online conferencing applications.
\newblock In \emph{Interspeech}, 2022.
\newblock URL \url{https://api.semanticscholar.org/CorpusID:247794123}.

\bibitem[Yoshimura et~al.(2016)Yoshimura, Henter, Watts, Wester, Yamagishi, and
  Tokuda]{yoshimura16_interspeech}
T.~Yoshimura, G.~E. Henter, O.~Watts, M.~Wester, J.~Yamagishi, and K.~Tokuda.
\newblock A hierarchical predictor of synthetic speech naturalness using neural
  networks.
\newblock In \emph{Interspeech 2016}, pages 342--346, 2016.
\newblock \doi{10.21437/Interspeech.2016-847}.

\bibitem[Yu et~al.(2024)Yu, Tang, Sun, Chen, Tan, Li, Lu, Ma, and
  Zhang]{yu2024connecting}
W.~Yu, C.~Tang, G.~Sun, X.~Chen, T.~Tan, W.~Li, L.~Lu, Z.~Ma, and C.~Zhang.
\newblock Connecting speech encoder and large language model for asr.
\newblock In \emph{ICASSP 2024-2024 IEEE International Conference on Acoustics,
  Speech and Signal Processing (ICASSP)}, pages 12637--12641. IEEE, 2024.

\bibitem[Zezario et~al.(2025)Zezario, Siniscalchi, Wang, and
  Tsao]{zezario2025study}
R.~E. Zezario, S.~M. Siniscalchi, H.-M. Wang, and Y.~Tsao.
\newblock A study on zero-shot non-intrusive speech assessment using large
  language models.
\newblock In \emph{ICASSP 2025-2025 IEEE International Conference on Acoustics,
  Speech and Signal Processing (ICASSP)}, pages 1--5. IEEE, 2025.

\bibitem[Zheng et~al.(2023)Zheng, Chiang, Sheng, Zhuang, Wu, Zhuang, Lin, Li,
  Li, Xing, Zhang, Gonzalez, and Stoica]{vicuna}
L.~Zheng, W.-L. Chiang, Y.~Sheng, S.~Zhuang, Z.~Wu, Y.~Zhuang, Z.~Lin, Z.~Li,
  D.~Li, E.~P. Xing, H.~Zhang, J.~E. Gonzalez, and I.~Stoica.
\newblock Judging llm-as-a-judge with mt-bench and chatbot arena.
\newblock In \emph{Proceedings of the 37th International Conference on Neural
  Information Processing Systems}, NIPS '23, Red Hook, NY, USA, 2023. Curran
  Associates Inc.

\bibitem[Zheng et~al.(2024)Zheng, Peng, Ma, Chen, Choi, and
  Harwath]{zheng2024bat}
Z.~Zheng, P.~Peng, Z.~Ma, X.~Chen, E.~Choi, and D.~Harwath.
\newblock Bat: Learning to reason about spatial sounds with large language
  models.
\newblock \emph{arXiv preprint arXiv:2402.01591}, 2024.

\end{thebibliography}

\end{document}